\documentclass[pra,twocolumn,showpacs]{revtex4}
\usepackage[dvips]{graphicx}
\input{epsf}
\usepackage{amsmath}
\usepackage{amssymb}
\usepackage{stmaryrd}
\usepackage[mathscr]{eucal}

\newcommand{\beq}{\begin{equation}}
\newcommand{\eeq}{\end{equation}}
\newcommand{\beqa}{\begin{eqnarray}}
\newcommand{\eeqa}{\end{eqnarray}}
\newcommand{\beqan}{\begin{eqnarray*}}
\newcommand{\eeqan}{\end{eqnarray*}}

\newcommand{\ket}[1]{| #1 \rangle}

\newcommand{\ox}{\otimes}

\newcommand{\proof}{\noindent {\bf Proof. }}
\newcommand{\qed}{\hfill $\Box$ \vskip 2ex}

\newtheorem{theorem}{Theorem}

\newtheorem{proposition}{Proposition}

\newtheorem{corollary}{Corollary}

\begin{document}

\title{Commuting multiparty quantum observables and local compatibility}
\author{Claudio Altafini}
\affiliation{SISSA-ISAS  \\
International School for Advanced Studies \\
via Beirut 2-4, 34014 Trieste, Italy }

\pacs{03.65.Ta, 03.65.Ud, 03.67.-a}


\begin{abstract}
A formula for the commutator of tensor product matrices is used to shows that, for qubits, compatibility of quantum multiparty observables almost never implies local compatibility at each site and to predict when this happens/does not happen in a concise manner.
In particular, it is shown that two ``fully nontrivial'' $n$-qubit observables are compatible locally and globally if and only if they are equal up to sign.
In addition, the formula gives insight into the construction of new paradoxes of the type of the Kochen-Specker Theorem, which can then be easily rephrased into proposals for new no hidden variable experiments of the type of the  ``Bell Theorem without inequalities''.

\end{abstract}

\maketitle 



\section{Introduction}

In the literature that deals with Bell theorem and its consequences for quantum mechanics of multiparty systems, see \cite{Bell1,Kafatos1,Greenberger1,Mermin1,Peres2,Jordan1,Bertlmann1,Werner2} for an overview, a common aspect of the tests aiming at proving/disproving the notion of local realism is the use of two or more measurements along suitably chosen observables.
This unavoidably evokes the notion of {\em compatibility} of observables, which is one of the cornerstones of quantum mechanics and which affirms that simultaneous measurements are possible for observables that commute, since they share common eigenstates.
When we have a multiparty quantum state, such a compatibility condition is naturally referred to the observables of the compound state, at least as long as we regard the multiparty system as one wavefunction.
It is known that {\em global compatibility} (intended as compatibility among the observables of the compound system) not always corresponds to compatibility of the observables of each of the (possibly far apart) parties and that this lack of {\em local compatibility} in its turn may induce inconsistencies and violations of local hidden variable models, but there does not seem to be much awareness of how the two compatibility conditions are related.
Scope of this work is to make the connection explicit in the case of qubits and to start a systematic investigation of the consequences of neglected local incompatibilities. 
In order to do that, we use a formula for the commutator of multiparty observables in terms of commutator/anticommutators operations at each site provided in \cite{Cla-spin-tens1} and reproduced in the Appendix.

The ``qualitative'' paradoxes obtainable by means of the ``Kochen-Specker (KS) Theorem'' (\cite{Mermin1,Peres2}) which are known to be due to the ``counterfactual logic'' following from the simultaneous application of multiple globally compatible observables rather than to the entanglement properties of a state \cite{Mermin1,Peres2,Jordan1}, can then be reformulated in terms of hidden noncommuting local observables and studied by looking at the corresponding product of observables at each site.
We will see how it is possible to obtain an entire class of such KS no-hidden variable results for $n$ qubits in the case of $n$ odd, and how, in their turn, they induce a class of paradoxes of the type of the ``Bell Theorem without inequalities'' on suitably chosen entangled states, of which the GHZ case \cite{Greenberger1} is just one example.

A natural way to avoid inconsistencies of the KS type is to require that globally commuting observables also commute at each site.
Just like for a single qubit two nontrivial observables commute if and only if they are dichotomic, for $n$ qubits it will be shown that, provided we restrict to {\em fully nontrivial} observables (i.e., that do not act trivially on any of the parties), local and global compatibility hold simultaneously only for observables that are identical up to the sign (i.e., dichotomic). 
If instead we require only global compatibility, then it turns out that there is always an even number of hidden local nonzero commutations.
For a single qubit, our observables are ``complementary'' in the sense of \cite{Brukner1}, namely when one measurement is well-defined the other will be maximally uncertain.
For $n$ qubits, our result translates then into the following: two compatible fully nontrivial observables are always corresponding locally to a pair of complementary measurements in an even number of parties, and this number is zero only when the two multiparty observables are dichotomic.

\section{2-qubit case}

Consider the complete set of orthogonal observables obtained by taking tensor products of the Pauli matrices $ \sigma_j $, $ j \in \{ 0, \ldots , 3\} $.
For an $n$-qubit state, an observable will be denoted as $ \sigma_{j_1\ldots j_n } = \sigma_{j_1} \ox \ldots\ox \sigma_{j_n} $ \footnote{All results could be reformulated in terms of observables $ A= a^j \sigma_j $, $ j=0,\ldots, 3$, and tensor products of that, but we find it more complicated to be ``exhaustive'' in this way than with orthogonal basis elements.}.
Given the 1-qubit observables $ \sigma_j $ and $ \sigma_k $,  $ j, \, k \in \{ 0, \ldots , 3 \}  $, for their commutator and anticommutator we have, respectively,
\beqa
[ \sigma_{j} , \, \sigma_{k} ] &  \neq 0  \quad & \text{for $ (j, k ) = (1,2), (1,3), (2,3) $}
\label{eq:comm1} \\
&& \text{   and permutations, }   \nonumber
\eeqa
\beqa
\{ \sigma_{j} , \, \sigma_{k} \} & \neq 0 \quad  & \text{for $ (j, k ) = (0,1), (0,2), (0,3)$} ,
 \label{eq:anticomm1} \\
&& \text{ $\qquad \; \, (0,0), (1,1), (2,2), (3,3) $}   
\nonumber \\
 && \text{ and permutations, }  \nonumber
\eeqa
and 0 otherwise in both \eqref{eq:comm1} and \eqref{eq:anticomm1}.
Looking at \eqref{eq:comm1}-\eqref{eq:anticomm1} it is easy to realize that for any pair of observables $ \sigma_j $, $ \sigma_k$, if one excludes the trivial cases (i.e., when $j$ and/or $k$ are $0$) the only possibility for $ \sigma_j $ and $ \sigma_k $ to commute is that $ \sigma_j = \pm \sigma_k$, i.e., the  observables are identical up to the sign, as is well-known.

\begin{proposition}
\label{prop:no-both}
There is no pair $ (j,k) $ such that both $ [ \sigma_{j} , \, \sigma_{k} ] \neq 0 $ and $ \{ \sigma_{j} , \, \sigma_{k} \} \neq 0 $, or such that both $ [ \sigma_{j} , \, \sigma_{k} ] = 0 $ and $ \{ \sigma_{j} , \, \sigma_{k} \} = 0 $ \footnote{In terms of ``general'' observables $ A= a^0 \sigma_0 + \vec{a} \cdot \vec{\lambda} $ and $  B= b^0 \sigma_0 + \vec{b} \cdot \vec{\lambda} $, Proposition~\ref{prop:no-both} simply means that $ A$ and $B$ either commute (and in this case $  \vec{a} \parallel \vec{b} $ and $ \{ A, B \} \neq 0 $) or anticommute ($  \vec{a} \nparallel \vec{b} $ and $[ A, B ]\neq 0 $).}.
\end{proposition}
\proof
Given a pair of indexes $ (j, k ) $, \eqref{eq:comm1}-\eqref{eq:anticomm1} exhaust all their possible combinations.
Hence one and only one between $ [ \sigma_{j} , \, \sigma_{k} ] $ and $ \{ \sigma_{j} , \, \sigma_{k} \} $ must be nonzero.
\qed

There is a simple way to check whether two quantum observables commute: since both are Hermitian matrices, the answer is uniquely given by their matrix commutator.
If the two matrices are tensor product of matrices of compatible dimensions, then the matrix commutator can be ``destructured'' into sums of tensor products of two basic building blocks: commutators and anticommutators of one-party matrices, see the Appendix. 
For 2-qubit observables $ \sigma_{jk} = \sigma_j \ox \sigma_k $, this becomes:
\beq
[\sigma_{jk}, \, \sigma_{lm} ] = \frac{1}{2} \left( [ \sigma_j , \, \sigma_l ] \otimes  \{  \sigma_k , \, \sigma_m \} + \{  \sigma_j , \, \sigma_l \} \otimes  [ \sigma_k, \, \sigma_m ] \right) .
\label{eq:comm-2-qubit}
\eeq

\begin{proposition}
\label{prop:comm-anti-2-qubit1}
Exactly 2 of the 4 commutator/anticommutator operations of \eqref{eq:comm-2-qubit} must be $ \neq 0 $.
Furthermore, at most one of the terms in \eqref{eq:comm-2-qubit} can be nonzero.
\end{proposition}
\proof
The proof of the first part follows straightforwardly from Proposition~\ref{prop:no-both}.
Concerning the second part, assume the first term is nonzero.
Then the pair $ (j,l)$ must be one of \eqref{eq:comm1} and $(k,m) $ one of \eqref{eq:anticomm1}. Therefore, from Proposition~\ref{prop:no-both}, $ \{ \sigma_{j} , \, \sigma_{l} \} =0 $ and $ [ \sigma_{k} , \, \sigma_{m} ] =0$.
\qed

There are 2 and only 2 possible combinations leading to a commutation in \eqref{eq:comm-2-qubit}.
\begin{proposition}
\label{prop:comm-anti-2-qubit3}
$ [\sigma_{jk}, \, \sigma_{lm} ] = 0 $ if and only if one of the two possibilities below is verified:
\begin{description}
\item[2.i  ]  $  [ \sigma_{j} , \, \sigma_{l} ] = [ \sigma_{k} , \, \sigma_{m} ]=0 $,
\item[2.ii ]  $  \{ \sigma_{j} , \, \sigma_{l} \} = \{ \sigma_{k} , \, \sigma_{m} \}=0 $.
\end{description}
\end{proposition}

\proof
Follows from Propositions~\ref{prop:no-both}~and~\ref{prop:comm-anti-2-qubit1}. In order for $ \sigma_{jk} $ and $ \sigma_{lm} $ to commute, one operation on each summand of \eqref{eq:comm-2-qubit} must be null and one nonnull.
\qed
Proposition~\ref{prop:comm-anti-2-qubit1} obviously implies that the globally commuting observables of Proposition~\ref{prop:comm-anti-2-qubit3} are such that
\begin{description}
\item[2.i  ]  $  \{ \sigma_{j} , \, \sigma_{l} \} \neq 0 $, $ \{ \sigma_{k} , \, \sigma_{m} \}\neq 0 $,
\item[2.ii ]  $  [ \sigma_{j} , \, \sigma_{l} ] \neq 0 $, $  [ \sigma_{k} , \, \sigma_{m} ]\neq 0 $,
\end{description}
respectively \footnote{For $ A = A_1 \ox A_2 = a^j_1 \sigma_j \ox a^j_2 \sigma_j  $ and $ B = B_1 \ox B_2 = b^j_1 \sigma_j \ox b^j_2 \sigma_j $, Proposition~\ref{prop:comm-anti-2-qubit3} translates into $ [ A, B]=0 $ if and only if one of the two holds: (i) $ [A_1, B_1 ] = [A_2, B_2]=0 $ ($\vec{a}_k \parallel \vec{b}_k$, $ k=1,2$, implying that $ \{ A_k , B_k \} \neq 0 $); (ii) $ \{ A_1, B_1 \} = \{ A_2, B_2\} =0 $ ($\vec{a}_k \nparallel \vec{b}_k$, $ k=1,2$, hence $ [ A_k , B_k ] \neq 0 $).}.
Notice that whenever $  [ \sigma_{jk} , \, \sigma_{lm} ] = 0$ the anticommutator  
\[
\{ \sigma_{jk} , \, \sigma_{lm} \} =  \frac{1}{2} \left( [ \sigma_j , \, \sigma_l ] \otimes  [  \sigma_k , \, \sigma_m ] + \{  \sigma_j , \, \sigma_l \} \otimes  \{ \sigma_k, \, \sigma_m \} \right)  
\]
is nonzero and viceversa.
Hence one always has
\beq
 \sigma_{jk}  \sigma_{lm} = \frac{1}{2} \left( [ \sigma_{jk} , \, \sigma_{lm} ] + \{ \sigma_{jk} , \, \sigma_{lm} \}  \right)\neq 0  .
\label{eq:prod-2-qubit}
\eeq
While in case {\bf 2.ii} a couple of observables that commute at global level is actually  hiding local noncommutativity on both sites, in case {\bf 2.i} local and global compatibility coexist.
Excluding the 1-qubit operators ($\sigma_{j0} $ and $ \sigma_{0k}$), it is easily seen that {\bf 2.i} requires that two observables $ \sigma_{jk} $ and $ \sigma_{lm} $ have $ j=l $ and $ k=m $, i.e., $ \sigma_{jk} = \pm \sigma_{lm} $ dichotomic observables, just like for the single qubit case.

The hidden noncompatibilities of Case {\bf 2.ii} are indeed the source of algebraic contradictions leading to violations of local hidden variable models for instance in the form given in Mermin \cite{Mermin1} based on 3 globally commuting observables (see also Ch.~7.1 of \cite{Peres2}) and not involving statistical correlations of ensembles.
Consider the last row and column of the example in Fig.~3 in \cite{Mermin1} involving the following two triples of mutually globally commuting observables: $ \sigma_{12}$, $ \sigma_{21}$,  $ \sigma_{33}$ and $ \sigma_{11}$, $ \sigma_{22}$, $ \sigma_{33} $.
All pairwise commutation relations are of the type {\bf 2.ii}.
Looking at what happens to the single slot commutators of each (spatially separated) party, for the first qubit there is  no difference: they are $ [ \sigma_{1} , \, \sigma_{2} ] $ $ [ \sigma_{1} , \, \sigma_{3} ] $ and $ [ \sigma_{2} , \, \sigma_{3} ] $ for both triples.
Looking at the second slot instead we have $ [ \sigma_{1} , \, \sigma_{3} ] $ and $ [ \sigma_{2} , \, \sigma_{3} ] $ in common but the third one is  $ [ \sigma_{2} , \, \sigma_{1} ] $ for the first triple and  $ [ \sigma_{1} , \, \sigma_{2} ] $ for the second one.
Hence the sign difference leading to the violation of the local hidden variables model.
More formally, what we are using is the following formula valid for any triple of mutually globally commuting 2-qubit observables:
\beq
\begin{split}
&  \sigma_{jk} \sigma_{lm} \sigma_{rs}  = \frac{1}{4} \{  \{ \sigma_{jk} , \sigma_{lm} \} ,  \sigma_{rs} \}   \\
& = \frac{1}{16} \left( 
[ [ \sigma_j , \sigma_l ], \sigma_r]
\ox [ [ \sigma_k , \sigma_m ], \sigma_s] 
\right. \\ & \quad  \left. 
+ \{ [ \sigma_j , \sigma_l ], \sigma_r \} 
\ox \{ [ \sigma_k , \sigma_m ], \sigma_s\}
\right. \\ & \quad \left. 
+[ \{ \sigma_j , \sigma_l \}, \sigma_r]
\ox [ \{ \sigma_k , \sigma_m \}, \sigma_s] 
\right. \\ &\quad  \left. 
+ \{ \{ \sigma_j , \sigma_l \}, \sigma_r\} 
\ox \{ \{ \sigma_k , \sigma_m \}, \sigma_s\} \right) .
\end{split}
\label{eq:triple-prod-comm}
\eeq
Owing to the non-disturbing nature of such measurements, nothing forbids to think of the sequence as applied simultaneously (yielding a {\em product of observables} as in \eqref{eq:triple-prod-comm}) and of the corresponding joint probability being observed.
In \eqref{eq:triple-prod-comm}, since both triples of commuting observables belong (pairwise) to case {\bf 2.ii} above, they must obey $ \sigma_j \neq \sigma_l \neq \sigma_r $ and $ \sigma_k \neq \sigma_m \neq \sigma_s $.
From the basic commutation relations, $ [ \sigma_j , \sigma_l ] = \pm  i \sigma_r $ and $ [ \sigma_k , \sigma_m ] = \pm i \sigma_s $. 
Therefore on each site only the second term of \eqref{eq:triple-prod-comm} is nonzero and $ \sigma_{jk} \sigma_{lm} \sigma_{rs} = \pm\frac{1}{4} \sigma_0 \ox \sigma_0   $.
Hence $ \sigma_{jk} \sigma_{lm} \sigma_{rs} \ket{\psi} = \pm\frac{1}{4} \sigma_{00}\ket{\psi} =  \pm\frac{1}{4} \ket{\psi} $ for any $ \ket{\psi}$.
The sign varies according to the choice of indexes and for the two triples mentioned above one gets two opposite signs regardless of the value of the wavefunction: 
\beq
\sigma_{12} \sigma_{21} \sigma_{33} = -\sigma_{11} \sigma_{22} \sigma_{33} .
\label{eq:triple-meas-2}
\eeq
The inconsistency of \eqref{eq:triple-meas-2} can be rephrased as follows.
Assuming local realism holds, each triple of measurements locally consists of a cascade of measurements along $ \sigma_1 $, $ \sigma_2 $ and $ \sigma_3 $, in different orders.
Globally, the compatibility of the 3 observables guarantees that the order is irrelevant.
However, with the two triples above for the second party we have that the two ordering $ \sigma_2  \sigma_1 \sigma_3 $ and $ \sigma_1  \sigma_2 \sigma_3 $ yield opposite signs (as expected from a local point of view because of local noncompatibility and ``complementarity'').
Hence local and global points of view are in conflict.
Notice from \eqref{eq:triple-meas-2} how the paradox holds also for two pairs of commuting observables.
If we drop $ \sigma_{33} $ in \eqref{eq:triple-meas-2}, however, the inconsistency arguments become a function of $ \ket{\psi} $ (verified almost always but not always) because the product of two observables is not a constant as in \eqref{eq:triple-meas-2}.
Notice further the lack of joint (global) compatibility between the two sets of observables.

\section{3-qubit case}

The extension to 3 qubits can be analyzed by similar methods.
For example, the commutator is (see \eqref{eq:app-Lieb2})
\beq
\begin{split}
& [ \sigma_{jkl} , \, \sigma_{mpq} ]  = \frac{1}{4} \left( 
[\sigma_j,  \sigma_m ]\ox \{\sigma_k ,   \sigma_p \} \ox  \{ \sigma_l, \sigma_q \}
\right. \\
& \qquad  \qquad \qquad \left. 
+ \{\sigma_j, \sigma_m \}\ox [ \sigma_k , \sigma_p ] \ox \{\sigma_l , \sigma_q \} 
\right. \\
& \qquad \qquad \qquad\left. 
+ \{\sigma_j , \sigma_m \} \ox \{ \sigma_k , \sigma_p \} \ox [ \sigma_l , \sigma_q ] 
\right. \\
& \qquad \qquad \qquad \left. 
+ [ \sigma_j, \sigma_m ]\ox [ \sigma_k , \sigma_p ] \ox  [ \sigma_l , \sigma_q ] \right) .
\end{split}
\label{eq:comm-3-qubit}
\eeq
Since (see \eqref{eq:app-antic2})
\beq
\begin{split}
& \{ \sigma_{jkl} , \, \sigma_{mpq} \}  = \frac{1}{4} \left( 
[\sigma_j,  \sigma_m ]\ox [\sigma_k ,   \sigma_p ] \ox  \{ \sigma_l, \sigma_q \}
\right. \\
& \qquad  \qquad \qquad \;\; \left. 
+ [\sigma_j, \sigma_m ]\ox \{ \sigma_k , \sigma_p \} \ox [\sigma_l , \sigma_q ] 
\right. \\
& \qquad \qquad \qquad \; \; \left. 
+ \{\sigma_j , \sigma_m \} \ox [ \sigma_k , \sigma_p ] \ox [ \sigma_l , \sigma_q ] 
\right. \\
& \qquad \qquad \qquad \;\;  \left. 
+ \{ \sigma_j, \sigma_m \}\ox \{ \sigma_k , \sigma_p \} \ox  \{ \sigma_l , \sigma_q \} \right) ,
\end{split}
\label{eq:anticomm-3-qubit}
\eeq
the product 
\beq
 \sigma_{jkl}  \sigma_{mpq}   = \frac{1}{2} \left(  [ \sigma_{jkl} , \, \sigma_{mpq} ] +  \{ \sigma_{jkl} , \, \sigma_{mpq} \} \right) \neq 0
\label{eq:prod-3-qubit}
\eeq
always.
From \eqref{eq:comm-3-qubit} and \eqref{eq:anticomm-3-qubit}, all 8 possible combinations of commutators and anticommutators are present in a product like \eqref{eq:prod-3-qubit}.
From Proposition~\ref{prop:no-both}, one and only one of the 8 summands is nonzero.
Furthermore, up to a (pairwise) permutation of the 3 indexes, there are 4 possible combinations for the commutators/anticommutators of \eqref{eq:comm-3-qubit}:
\begin{description}
\item[3.i] 
$\quad 
\begin{cases} 
 \{ \sigma_{j} , \, \sigma_{m} \} = 0 \\
 \{ \sigma_{k} , \, \sigma_{p} \} = 0 \\
 \{ \sigma_{l} , \, \sigma_{q} \} = 0 
\end{cases}
\quad \Rightarrow \quad 
\begin{cases} 
 [ \sigma_{j} , \, \sigma_{m} ] \neq 0 \\
 [ \sigma_{k} , \, \sigma_{p} ] \neq 0 \\
 [ \sigma_{l} , \, \sigma_{q} ] \neq 0 
\end{cases}
$
\item[3.ii] 
$\quad 
\begin{cases} 
 [ \sigma_{j} , \, \sigma_{m} ] = 0 \\
 \{ \sigma_{k} , \, \sigma_{p} \} = 0 \\
 \{ \sigma_{l} , \, \sigma_{q} \} = 0 
\end{cases}
\quad \Rightarrow \quad 
\begin{cases} 
 \{ \sigma_{j} , \, \sigma_{m} \} \neq 0 \\
 [ \sigma_{k} , \, \sigma_{p} ] \neq 0 \\
 [ \sigma_{l} , \, \sigma_{q} ] \neq 0 
\end{cases}
$
\item[3.iii] 
$\quad 
\begin{cases} 
 [ \sigma_{j} , \, \sigma_{m} ] = 0 \\
 [ \sigma_{k} , \, \sigma_{p} ] = 0 \\
 \{ \sigma_{l} , \, \sigma_{q} \} = 0 
\end{cases}
\quad \Rightarrow \quad 
\begin{cases} 
 \{ \sigma_{j} , \, \sigma_{m} \} \neq 0 \\
 \{ \sigma_{k} , \, \sigma_{p} \} \neq 0 \\
 [ \sigma_{l} , \, \sigma_{q} ] \neq 0 
\end{cases}
$
\item[3.iv] 
$\quad 
\begin{cases} 
 [ \sigma_{j} , \, \sigma_{m} ] = 0 \\
 [ \sigma_{k} , \, \sigma_{p} ] = 0 \\
 [ \sigma_{l} , \, \sigma_{q} ] = 0 
\end{cases}
\quad \Rightarrow \quad 
\begin{cases} 
 \{ \sigma_{j} , \, \sigma_{m} \} \neq 0 \\
 \{ \sigma_{k} , \, \sigma_{p} \} \neq 0 \\
 \{ \sigma_{l} , \, \sigma_{q} \} \neq 0 .
\end{cases}
$
\end{description}
\begin{proposition} 
$  [ \sigma_{jkl} , \, \sigma_{mpq} ] =0 $ if and only if we are in the cases {\bf 3.ii} and {\bf 3.iv}.
\end{proposition}
In fact, {\bf 3.i} implies that the fourth term of \eqref{eq:comm-3-qubit} is always nonzero and {\bf 3.iii} implies it is nonzero the third one.
Of the two combinations yielding global compatibility, only {\bf 3.ii} hides local noncommuting observables.
In this case \eqref{eq:comm-3-qubit} contains:
\begin{itemize}
\item one 1-qubit hidden commutation in 2 of the 4 terms (the second and the third);
\item two 1-qubit hidden commutations in one of the 4 terms (the fourth).
\end{itemize}

Using the expression \eqref{eq:prod-3-qubit} for the product of observables, it is easy to construct new paradoxes of the KS type involving exclusively counterfactual arguments among observables.
Consider the 5 mutually commuting 3-qubit observables
\beq
\sigma_{jjj}, \, \sigma_{jkk}, \, \sigma_{kjk}, \, \sigma_{kkj}, \, \sigma_{ll0}, \; \, j,k,l\in \{ 1,2,3\}, \; j\neq k\neq l.
\label{eq:5-comm-3-quibt}
\eeq
Unlike the 2-qubit case discussed above, the corresponding measurements are now, in principle, attainable by a single experimental apparatus as the 5 observables have a complete set of common eigenkets.
When we compute explicitly the following two triple products, we have a dichotomy:
\beq
 \sigma_{ll0}\sigma_{jjj}\sigma_{kkj} =  \sigma_{000} = - \sigma_{ll0}\sigma_{kjk}\sigma_{jkk} .
\label{eq:paradox-5-comm-3-qubit}
\eeq
Looking at what happens at each (spatially separated) party, \eqref{eq:paradox-5-comm-3-qubit} corresponds on the left hand side to measuring $ \sigma_l \sigma_j \sigma_k $ at the first and second site and $ \sigma_0 \sigma_j \sigma_j $ at the third one, and, on the right hand side, to $ \sigma_l \sigma_k \sigma_j $, $ \sigma_l \sigma_j \sigma_k $ and $ \sigma_0 \sigma_k \sigma_k $ respectively.
For the third party everything is commuting and $ \sigma_0 \sigma_j \sigma_j = \sigma_0 \sigma_k \sigma_k =  \sigma_0$.
The second observer applies the same ordered sequence of operators, only the first one has a difference in the two ordering. 
From this follows that local realism is falsified because globally the order is irrelevant by assumption, while locally it leads to the opposite signs in \eqref{eq:paradox-5-comm-3-qubit}.
Using \eqref{eq:anticomm-3-qubit}, the existence of the paradox is revealed by the odd difference of signs in the ``hidden'' local commutations for the following two multiplications:
\beqa
\sigma_{jjj}\sigma_{kkj} & = & 
 \frac{1}{2} 
[ \sigma_j , \sigma_k ] \ox 
[  \sigma_j , \sigma_k ] \ox 
\{  \sigma_j , \sigma_j \} 
\label{eq:paradox-5-comm-3-qubit-a} \\
\sigma_{kjk}\sigma_{jkk} & = & 
 \frac{1}{2} 
[ \sigma_k , \sigma_j ] \ox 
[  \sigma_j , \sigma_k ] \ox 
\{  \sigma_k , \sigma_k \} .
\label{eq:paradox-5-comm-3-qubit-b}
\eeqa
While the KS theorem yields a logical, state independent contradiction, it is possible to transform it into an instance of the Bell Theorem without inequalities i.e., in a paradox expressible in terms of a suitably chosen entangled state. 
The situation described above includes as special case (considering only the first 4 observables of \eqref{eq:5-comm-3-quibt} with $j=1$ and $ k=2 $) the GHZ example \cite{Greenberger1}, dealing with the state $ \ket{\psi_1} =\frac{1}{\sqrt{2} } \left( \ket{000} + \ket{111} \right) $.
Varying the indexes $ j,k,l$, one obtains a number of alternative no hidden variable tests for different entangled states. For example:
\begin{itemize}
\item $ j=2$, $ k=1$ (and $ l=3$): 
\[
 \ket{\psi_2}= \frac{1}{\sqrt{2} } \left( \ket{000} -i  \ket{111} \right) ;
\]
\item $ j=3$, $ k=1$ (and $ l=2$): 
\[ \ket{\psi_3}= \frac{1}{2 } \left( \ket{000} -  \ket{011} - \ket{101} - \ket{110}  \right) ;
\]
\item $ j=2$, $ k=3$ (and $ l=1$): 
\beqan 
\ket{\psi_4} & = & \frac{1}{2\sqrt{2} } \left( \ket{000} + \ket{011} + \ket{101} +\ket{110}  \right) \\
&&  -\frac{i}{2\sqrt{2} } \left( \ket{001} + \ket{010} + \ket{100} +\ket{111}  \right)  .
\eeqan
\end{itemize}
It is straightforward to verify that all of these states have bipartite entanglement between each pair of qubits.
All of the choices of $ j$, $k,$ and $ l$ lead to a potential experimental test of local hidden variable violation, alternative to the standard GHZ setting of \cite{Greenberger1}.
Consider for example $ \ket{\psi_3}$. 
It is straightforward to show that $ \ket{\psi_3} $ is an eigenstate of the 4 detectors $ \sigma_{113}$, $ \sigma_{131}$, $ \sigma_{311}$, $ \sigma_{333}$ and that 
\beq
\sigma_{113}\ket{\psi_3} = \sigma_{131}\ket{\psi_3} = \sigma_{311}\ket{\psi_3} =- \ket{\psi_3} 
\label{eq:GHZ-alternat1}
\eeq
while 
\beq
 \sigma_{333}\ket{\psi_3} = \ket{\psi_3} .
\label{eq:GHZ-alternat2}
\eeq
This combination is not compatible with any assignment of local elements of reality, just like in the GHZ case.

\section{$n$-qubit case}

The considerations above extend to a generic number $n$ of qubits.
An observable $ \sigma_{j_1\ldots j_n }$ will be denoted fully nontrivial when $ j_p \neq 0 $ $ \forall \, p=1, \ldots, n $.
Of the various cases of compatible fully nontrivial observables arising for $n$ qubits, only one will ensure local and global compatibility.

\begin{theorem}
\label{thm:n-qubit1}
Two fully nontrivial $n$-qubit observables are locally and globally compatible if and only if they are equal up to sign.
If instead they are only globally compatible, then they are always locally noncommuting in an even number of sites.
\end{theorem}
\proof
Call $ \sigma_{j_1 \ldots j_n }$ and $ \sigma_{k_1 \ldots k_n} $ the two observables.
They are dichotomic, $   \sigma_{j_1 \ldots j_n } = \pm  \sigma_{k_1 \ldots k_n} $, if and only if 
\beq
[ \sigma_{j_p} , \, \sigma_{k_p} ] =0 
\label{eq:proof-comm-1qub}
\eeq
$ \forall \, p=1, \ldots , n $, which corresponds to $ \sigma_{j_p} = \pm \sigma_{k_p} $ since they are fully nontrivial.
In the proof of the first part, one direction is obvious, the other will be shown by induction.
From above, the claim is true for $2,3$ qubits.
Assume it is true for the two $(n-1)$-party observables $ \sigma_{j_1 \ldots j_{n-1}} $ and $ \sigma_{k_1 \ldots k_{n-1}} $, i.e., \eqref{eq:proof-comm-1qub} holds up to $ p=n-1$, and $ [ \sigma_{j_1 \ldots j_{n-1}} , \, \sigma_{k_1 \ldots k_{n-1}} ] =0$.
Then it is enough to write the commutator as (see also the proof of Proposition 1 of \cite{Cla-spin-tens1}) 
\beq
\begin{split}
[ \sigma_{j_1 \ldots j_n } , \, \sigma_{k_1 \ldots k_n} ]  =  & \frac{1}{2} \left( [ \sigma_{j_1 \ldots j_{n-1}} , \, \sigma_{k_1 \ldots k_{n-1}} ] \ox \{  \sigma_{j_n} , \, \sigma_{k_n} \} \right. \\
& \left. + \, \{ \sigma_{j_1 \ldots j_{n-1}} , \, \sigma_{k_1 \ldots k_{n-1}} \} \ox [ \sigma_{j_n} , \, \sigma_{k_n} ] \right)
\end{split}
\label{eq:comm-n-qubit}
\eeq
which is zero if and only if $ [ \sigma_{j_n} , \, \sigma_{k_n} ] =0 $ (from $ \sigma_{j_1 \ldots j_{n-1}}  \sigma_{k_1 \ldots k_{n-1}} \neq 0 $, the $(n-1)$-qubit anticommutator must be nonzero).
Together with the induction assumption $  \sigma_{j_1 \ldots j_{n-1}} = \pm \sigma_{k_1 \ldots k_{n-1}} $, this yields the first claim.
Concerning the second one, also use induction, but on the twofold assumption (true for $ n-1 =2, 3 $): 
\begin{enumerate} 
\item $ [ \sigma_{j_1 \ldots j_{n-1}} , \, \sigma_{k_1 \ldots k_{n-1}} ] =0 $ with an even number of hidden nonzero commutations and $  [ \sigma_{j_n} , \, \sigma_{k_n} ] =0 $;
\item $ \{ \sigma_{j_1 \ldots j_{n-1}} , \, \sigma_{k_1 \ldots k_{n-1}} \} =0 $ with an odd number of hidden nonzero commutations (for 3 qubits see cases {\bf 3.i}  and {\bf 3.iii} above) and $  \{ \sigma_{j_n} , \, \sigma_{k_n} \} =0 $.
\end{enumerate}
In both cases the conclusion follows from \eqref{eq:comm-n-qubit}.
\qed

A straightforward consequence is the following. 
\begin{corollary}
\label{cor:even}
Compatible fully nontrivial observables must differ for an even number of indexes.
\end{corollary}

From the proof of Theorem~\ref{thm:n-qubit1}, it also follows that $  \sigma_{j_1 \ldots j_{n}} \sigma_{k_1 \ldots k_{n}} \neq 0 $ always and that one and only one of the $ 2^n $ terms in the product is nonzero.
When $  \sigma_{j_1 \ldots j_{n}} $ and $ \sigma_{k_1 \ldots k_{n}} $ are compatible, they admit simultaneous measurements.
Consider as before in correspondence of the simultaneous measurements the product of observables, rewritten in the form 
\beq
(\sigma_{j_1} \sigma_{k_1}) \ox  \ldots\ox ( \sigma_{j_n} \sigma_{ k_n} ) .
\label{eq:prod-2-n-qubits}
\eeq 
If $ j_1 \ldots j_{n} \neq k_1 \ldots k_{n} $, then in an even number of sites \eqref{eq:prod-2-n-qubits} corresponds to noncompatible, complementary experiments.

As in the 3-qubit case, new paradoxes of the KS type can be constructued using the product of observables \eqref{eq:prod-2-n-qubits}.
Consider the following set of $ n+1 $ mutually globally commuting observables (pairwise differing by 2 or 4 indexes)
\beq
\begin{split}
& \sigma_{j_1 \ldots j_n}, \, \; \sigma_{j_1 \ldots j_{n-2} k_{n-1} k_n}, \, \; \sigma_{j_1 \ldots j_{n-3} k_{n-2} k_{n-1} j_n }, \ldots ,\\
&  \sigma_{k_1 k_2 j_3 \ldots j_n }, \, \; \sigma_{k_1 j_2 \ldots j_{n-1} k_n }
\end{split}
\label{eq:n+1-comm-n-qubit}
\eeq
where $ k_p \neq j_p $.
The paradox is obtained by comparing $ \sigma_{j_1 \ldots j_n} $ with the product of the last $ n-1 $ observables, which can be rewritten as 
\beq
\begin{split}
& (\sigma_{j_1} \ldots \sigma_{j_1} \sigma_{k_1} \sigma_{k_1}) 
\ox (\sigma_{j_2} \ldots \sigma_{j_2} \sigma_{k_2} \sigma_{k_2} \sigma_{j_2})
\ox \ldots \\
& \ox (\sigma_{k_{n-1}} \sigma_{k_{n-1}} \sigma_{j_{n-1}}  \ldots \sigma_{j_{n-1}}) 
\ox (\sigma_{k_n} \sigma_{j_n}  \ldots  \sigma_{j_n} \sigma_{k_n}).
\end{split}
\label{eq:prod-n-comm-n-qubit}
\eeq
Assume $n$ odd.
Since $ \sigma_{k_p}\sigma_{k_p} = \sigma_0 $, then at the $p$-th site the products yield $  \sigma_{j_p} $ except for the last one, which gives $ - \sigma_{j_n} $.
Hence $   \sigma_{j_1 \ldots j_n} $ and \eqref{eq:prod-n-comm-n-qubit} differ only by a sign and we have a KS paradox as no hidden variable theory is compatible with such an assignment.
Clearly, in a pair of products of globally commuting observables counterfactual arguments occur whenever the difference between the two sequences consists of an odd number of permutations of local observables in one site, not balanced by another odd number of local permutations at any other site \footnote{If we use ``general'' observables $A$, $B$, $C$, etc., pointing out the occurrence of an inconsistency only by counting permutations becomes in general impossible, as the violations are no longer deterministic.}.
Attaching a suitable entangled state, one gets new instances of the Bell Theorem without inequalities.
For example, choosing $ j_1 = \ldots j_n =1 $ and $ k_1 = \ldots k_n =2 $, the standard $ n$-partite GHZ paradox is obtained for the state $ \ket{\phi_1} =\frac{1}{\sqrt{2} } \left( \ket{0\ldots 0} + \ket{1\ldots 1} \right) $.
Different choices of indexes yield hidden variables tests for different entangled states.
A few simple examples, dealing with families of Dicke states are:
\begin{itemize}
\item $ j_1 = \ldots j_n =3 $ and $ k_1 = \ldots k_n =1 $
\[ 
\ket{\phi_2} = \frac{1}{n-1} \sum_{m=0,2, \ldots, n-1} (-1)^{m/2} P ( \ket{n-m, m} ) 
\]
\item $ j_1 = \ldots j_n =3 $ and $ k_1 = \ldots k_n =1 $
\[ 
\begin{split}
\ket{\phi_3} = & \frac{1}{(n-1) \sqrt{2} }\left(  \sum_{m=0,2, \ldots, n-1}  P ( \ket{n-m, m} ) \right. \\
 & \left.  + (-1)^{(n-1)/2}\,  i\! \sum_{m=1,3, \ldots, n}  P ( \ket{n-m, m} ) \right)
\end{split} 
\]
\end{itemize}
where $ \ket{n-m, m} $ means $ n-m $ times spin down and $ m$ times up and $ P(\cdot ) $ means sum over all possible permutations of the $n$ spins.
More complicated states are obtained when $  j_1, \ldots, j_n $ (and/or $ k_1 , \ldots , k_n $) are not all equal.

Consider for example $ \ket{\phi_2}$. 
It is just a matter of recursive computation to show that $ \ket{\phi_2} $ is an eigenstate of the $ n+1 $ observables \eqref{eq:n+1-comm-n-qubit} and that 
\[
 \sigma_{3\ldots 311}\ket{\phi_2} =  \sigma_{3\ldots 3113}\ket{\phi_2} =\ldots = \sigma_{13 \ldots 31} \ket{\phi_2} =- \ket{\phi_2} 
\label{eq:GHZ-alternatn1}
\]
\[
\text{while} \qquad \qquad \qquad
 \sigma_{3 \ldots 3}\ket{\phi_2} = \ket{\phi_2} . \qquad \qquad \qquad \qquad \qquad
\label{eq:GHZ-alternatn2}
\]
implying incompatibility with any assignment of local elements of reality.
The $n+1 $ observables of \eqref{eq:n+1-comm-n-qubit} differ pairwise by 2 or 4 indexes.
Other families of mutually commuting observables differing for other even numbers of indexes lead to similar conclusions (one such family is used in \cite{Mermin2} to obtain the so-called Mermin-Klyshko inequality).
The ``parity'' conditions of Theorem~\ref{thm:n-qubit1} and Corollary~\ref{cor:even}, as well as the construction \eqref{eq:prod-n-comm-n-qubit}, seem to indicate that for $ n $ even all paradoxes are actually involving effectively only $ n-1 $ parties.

Notice that starting with 3 qubits, it is possible to have a wider variety of observables commuting locally and globally (other than just dichotomic), provided that one considers also 2-party ``nonlocal'' observables i.e., observables that violate the full nontriviality assumption.
In fact, \eqref{eq:proof-comm-1qub} is satisfied also when $ j_p $ or $ k_p $ are $0$.
For example, $ \sigma_{0kl}$, $ \sigma_{j0l}$, $\sigma_{jk0}$ and $\sigma_{jkl}$ are globally commuting observables leaving no hidden complementarity behind, as they are mutually of type {\bf 3.iv} in the classification above.
Since local and global compatibility coexist, these observables do not seem to bear any intrinsic contradiction {\em per se}.

\section{Conclusion}

If for multiparticle systems one focuses as in this work on what a global point of view of a multiple measurement is neglecting in terms of multiple local measurements, the result is rather disorienting: a local observer only aware of his side of the measurement process may be (almost always for qubits) induced to consider the cascade of measurements as ill-posed because of the noncommutativity of the reduced observables, while from the global perspective everything was set up according to the compatibility rules of quantum mechanics.
Decomposing the compatibility condition in terms of local commutators/anticommutators of each party makes the detection of such incompatibilities straightforward.

\appendix

\section{Commutators and anticommutators of tensor product matrices}
\label{app:lie-brack}
Given $ A_1, \ldots, A_n, B_1 , \ldots, B_n  \in M_{m} $, their commutator is (see \cite{Cla-spin-tens1} for a proof)
\beq
\begin{split}
& [ A_1\ox \ldots \ox A_n , \, B_1 \ox \ldots \ox B_n ]=  \\
& = \sum \frac{1}{2^{n-1}} \left( ( A_1, \, B_1 ) \ox ( A_2, \, B_2 ) \ox \ldots \ox  ( A_n, \, B_n ) \right) 
\end{split}
\label{n-fact-tens-Lieb}
\eeq
where in each summand the bracket $ ( \; \cdot \;, \; \cdot \; ) $ correspond $ k$ times, ($ k$ odd) to a commutator and $ n - k$ times to an anticommutator.
The sum is over all possible (nonrepeated) combinations of $ [ \; \cdot \;, \; \cdot \; ] $ and $  \{ \; \cdot \;, \; \cdot \; \} $, and over all odd $ k \in [1, \, n] $.
For $ n=2,3$ this corresponds to:
\beq
\begin{split}
& [ A_1 \otimes A_2  , \, B_1\otimes B_2 ]= \\ 
& =  \frac{1}{2} \left(  [ A_1, \, B_1 ] \otimes \{ A_2, \,  B_2 \} +  \{A_1, \,  B_1 \}  \otimes [ A_2, \, B_2 ] \right),
\end{split}
\label{eq:app-Lieb1}
\eeq
\beq
\begin{split}
& [ A_1 \otimes A_2\otimes A_3  , \, B_1\otimes B_2\otimes B_3 ]= \\
& =  \frac{1}{4}  \left( 
[A_1 , \,B_1 ] \otimes \{A_2 , \,B_2 \}\otimes \{ A_3, \, B_3\} 
\right. \\ & \left. \quad 
+\{ A_1, \,B_1 \} \otimes [A_2 , \,B_2 ]  \otimes \{A_3 , \,B_3 \}  
 \right. \\ & \left. \quad 
+\{ A_1, \,B_1 \}  \otimes \{A_2 , \,B_2 \}  \otimes [ A_3, \,B_3 ]
 \right. \\ & \left. \quad 
+ [A_1, \, B_1] \otimes[A_2 , \,B_2 ]  \otimes [ A_3, \,B_3 ] 
 \right),
\end{split}
\label{eq:app-Lieb2}
\eeq
If instead we take $ k$ even, the same formula \eqref{n-fact-tens-Lieb} gives the anticommutator of $ A_1, \ldots, A_n, B_1 , \ldots, B_n  \in M_{m} $.
For $ n=2,3$ one has:
\beq
\begin{split}
& \{ A_1 \otimes A_2  , \, B_1\otimes B_2 \} = \\ 
& =  \frac{1}{2} \left(  [ A_1, \, B_1 ] \otimes [ A_2, \,  B_2 ] +  \{A_1, \,  B_1 \}  \otimes \{ A_2, \, B_2 \} \right),
\end{split}
\label{eq:app-antic1}
\eeq
\beq
\begin{split}
& \{ A_1 \otimes A_2\otimes A_3  , \, B_1\otimes B_2\otimes B_3 \}= \\
& =  \frac{1}{4}  \left( 
[A_1 , \,B_1 ] \otimes [ A_2 , \,B_2 ] \otimes \{ A_3, \, B_3\} 
\right. \\ & \left. \quad 
+[ A_1, \,B_1 ] \otimes \{ A_2 , \,B_2 \}  \otimes [ A_3 , \,B_3 ]  
 \right. \\ & \left. \quad 
+\{ A_1, \,B_1 \}  \otimes [ A_2 , \,B_2 ]  \otimes [ A_3, \,B_3 ]
 \right. \\ & \left. \quad 
+ \{ A_1, \, B_1\} \otimes \{ A_2 , \,B_2 \}  \otimes \{ A_3, \,B_3 \} 
 \right),
\end{split}
\label{eq:app-antic2}
\eeq

\bibliographystyle{apsrev}
\small

\end{document}